\documentclass[conference]{IEEEtran}
\IEEEoverridecommandlockouts
\usepackage{cite}
\usepackage{amsmath,amssymb,amsfonts}
\usepackage{algorithmic}
\usepackage{graphicx}
\usepackage{textcomp}
\usepackage{xcolor}
\usepackage{hyperref}
\usepackage{booktabs}

\usepackage{glossaries}

\makeglossaries

\newacronym{mturk}{MTurk}{Amazon Mechanical Turk}

\newacronym{efps}{EFPS}{Effective Frames per second}
\newacronym{hit}{HIT}{Human intelligence task}
\newacronym{api}{API}{Application Programming Interface}

\newglossaryentry{maths}
{
    name=mathematics,
    description={Mathematics is what mathematicians do}
}

\def\BibTeX{{\rm B\kern-.05em{\sc i\kern-.025em b}\kern-.08em
    T\kern-.1667em\lower.7ex\hbox{E}\kern-.125emX}}
\begin{document}

\title{Evaluating hardware differences for crowdsourcing and traditional recruiting methods\\
\thanks{This work was supported by the EPSRC [EP/S023917/1].}
}

\author{\IEEEauthorblockN{Paul-David Joshua Zuercher}
\IEEEauthorblockA{\textit{Department of Engineering} \\
\textit{University of Cambridge}\\
Cambridge, United Kingdom \\
pdz20@cam.ac.uk}
}

\maketitle

\begin{abstract}
The most frequently used method to collect research data online is crowdsouring and its use continues to grow rapidly.
This report investigates for the first time whether researchers also have to expect significantly different hardware performance when deploying to \acrfull{mturk}.
This is assessed by collecting basic hardware parameters (Operating System, GPU, and used browser) from \gls{mturk} and a traditional recruitment method (i.e., snowballing). 
The significant hardware differences between crowdsourcing participants (\gls{mturk}) and snowball recruiting are reported including relevant descriptive statistics for assessing hardware performance of 3D web applications.
The report suggests that hardware differences need to be considered to obtain valid results if the designed experiment application requires graphical intense computations and relies on a coherent user experience of \gls{mturk} and more established recruitment strategies (i.e. snowballing).
\end{abstract}


\section{Introduction}
\paragraph{Motivation}
The most frequently used method to collect research data online is crowdsouring and its use continues to grow rapidly \cite{aguinis2021mturk}. 
Crowdsourcing is a form of recruitment in which a task requiring human intelligence is offered to an undefined network of people - also called "crowd" \cite{estelles2012towards}. 
Reasons for recruiting using crowdsourcing, through platforms such as \acrfull{mturk}, include better scalability, flexibility of research design, and affordability to perform experiments than other forms of recruiting \cite{aguinis2021mturk}. 
Besides these benefits, a major challenge of crowdsourcing research data is adapting the experiment to the crowdsourcing environment to obtain sound and valid research results \cite{aguinis2021mturk}. 

Hardware differences can impact experimental results that rely on a coherent user experience because better or worse user experience is impaired if the hardware is not performant enough \cite{chen2007review}.
For less hardware demanding 2D web applications, such as like surveys, potential hardware differences are less likely to affect the perceived user experience because almost all devices used should suffice the hardware requirements. 
For more complex 3D experiments applications (for example 3D game-like environments) the hardware performance requirements are higher, and significant differences in available hardware performance can lead to differences in user experience between the targeted group and the crowdsourcing group \cite{chen2007review}. 
Therefore, the investigated hardware differences are relevant to all studies that rely on a coherent user experience of crowdsourcing participants and more traditional forms of recruitment (i.e., snowballing) using 3D web applications, since significant hardware differences can lead to a significantly different user experience \cite{chen2007review}. 


\paragraph{Research question and hypothesis}
This report closes the research gap of missing hardware performance data by addressing the question: How hardware demanding can \acrfull{mturk} experiments be in comparison to traditional recruitment options? 
Since no prior research could be found that deemed the hardware differences significant, the null hypothesis is any difference in hardware performance between crowdsourcing and snowball recruitment is only due to chance. 
The proposed alternative hypothesis is that the hardware performance of \gls{mturk} is different to hardware performance of participants recruited using snowballing.

\paragraph{Choice of dataset}
This research question and hypothesis is answered based on an online survey of hardware components collated from (1) \gls{mturk} and (2) using snowballing (based on social media posts and by distributing over the research institute mailing list). 
The performance of the obtained qualitative data of GPU manufacturer and model is quantified in terms of performance using the benchmark results of \cite{userbenchmark} and compared for significant differences between the recruitment channels.

\section{Related Work}
In the following appropriate methods and techniques for (1) crowdsourcing, (2) analysing hardware performance, and (3) acquiring relevant hardware information are reviewed.

\subsection{Crowdsourcing with hardware demanding applications}
The biggest crowdsourcing platform is \acrfull{mturk}. 
On \gls{mturk} the main actors are the requesters and workers \cite{paolacci2010running}. 
Requesters pay a fee to create \acrfullpl{hit} which are defined as any task that is simple to solve with human intelligence \cite{aguinis2021mturk}. 
Workers can choose from and submit solutions for the list of \glspl{hit} \cite{paolacci2010running}. 
If the requester accepts the worker's contribution, the worker gets a fixed payment from the requester via \gls{mturk} \cite{paolacci2010running}. 
In the last decades, more and more researchers have used \gls{mturk} to perform experiments \cite{aguinis2021mturk}. 
Crowdsourcing experiments can follow common research guidelines for planning, implementing and reporting experiments to ensure the validity of crowdsourcing research \cite{aguinis2021mturk}. 

To the author's knowledge, no previous research has investigated the hardware performance demographic of \gls{mturk} workers. 
Despite not directly analysing and reporting hardware, previous approaches for managing hardware in crowdsourcing environments can be classified into two groups. 
Firstly, researchers created participant pools that suffice the application's hardware requirements by filtering out participants with insufficient hardware \cite{mohan2017crowdsourcing}.
Secondly, researchers reduced the application's hardware requirements and create coherent test environments by outsourcing demanding computations to an externally hosted server \cite{dolstra2013crowdsourcing, kan2014crowdadaptor, evans20143d}. 
In the second approach, the computation is separated into browser- and server-based rendering.
Thus, reducing the hardware requirements is achieved by outsourcing the computational intensive graphical operations to remote machines (servers) and whilst only rendering the resulting graphics in the browser which is locally less computational demanding \cite{evans20143d}.

The performance of browser-based rendering can be analysed based on the hardware and software used in the system, while remote rendering requires collecting additional information such as an sufficient connection between the local device to the remote device \cite{dolstra2013crowdsourcing, evans20143d}.
However, for both browser- and server-based rendering it is important to understand how demanding the local computations can be to ensure that (1) only allow participants with sufficient hardware performance and (2) operations exceeding the user's hardware capabilities are externalised to a sufficiently powerful server. 



\subsection{Analysing hardware performance}

In general, the applications execution depends on the system that operates the hardware \cite{paolacci2010running}. 
Since online crowdsourcing experiments using \gls{mturk} cannot legally ask participants to run any application outside of the web-browser, this report's analysis is restricted to web-applications. 
Web-applications are applications that run inside a web browser application on an operating system. 
Thus, the execution of a web-application is restricted by the (1) browser executing the web application, (2) operating system executing the web browser and (3) underlying computing hardware \cite{glander2013reweb3d}. 

Determining the expected web application performance can be done either by (a) inferring the performance capabilities based on collected hardware information (indirect) or (b) running benchmarks on a machine (direct) \cite{gries2004methods}. 
However, running these hardware performance tests directly can put a strain on the computer, takes time and moreover the results can suffer from high variance if the participant performs background tasks \cite{v2015build}. Therefore, it is more advisable to only collect the hardware data in remote experiments and infer the performance from the specifications indirectly.

After having obtained the hardware component information from the users, the hardware performance capability can be estimated.
The estimation of the hardware performance can either be done by analysing the technological specifications or by back-referencing application specific benchmark results \cite{paolacci2010running}.
Analysing the performance based on technological specifications is complex because small performance on one sub-component can reduce the overall systems' performance significantly depending on the application leading to many details that need to be investigated \cite{paolacci2010running}. Therefore, it is often preferable to determine the performance of a system based on benchmark results if the approximate application scenario is already known.

\subsection{Background literature for acquiring similar datasets.}

To survey the hardware information, one could (a) ask the participants to enter these information manually or (b) try to detect the information automatically. 
Letting the participants enter the information might lead to a self-selection bias since some of the participants might not know the information and the information would be subject to human error. The non-manual alternative, is to obtain the information using \acrfullpl{api} \cite{elbanna2018browsers}. 
Since \gls{mturk} only allows distributing web applications, information retrieval methods are restricted by the Web-API of the browsers \cite{elbanna2018browsers}. 
Research on collection of hardware information is of particular interest of privacy researchers in the area of "fingerprinting" \cite{elbanna2018browsers}. 
The objective of this research area is obtaining as much information as possible about a system to create a digital fingerprint of the system or user that allows identification or classification of the individual \cite{elbanna2018browsers}.
While the goal when collecting hardware information to estimate performance is not to identify the user, their methods allow obtaining (1) browser information, (2) operating system information, and (3) hardware information \cite{elbanna2018browsers}. 

\section{Method (design and implementation)}

The hardware performance is assessed by first collecting the hardware information of participants and secondly, assigning the participant's hardware to a performance score based on the selected benchmark.
Finally, appropriate tests from the area of statistical inference are used to evaluate the project's hypothesis.

\subsection{Experiment design}

The experiment uses two treatment groups to sample data using a web application. The web application handles the consent and hardware information collection from the participants.
The website is shared via two different links to differentiate data from crowdsourcing and traditional recruitment participants.
The crowdsourcing link is distributed as a \gls{hit} on \gls{mturk} and a request to participate on the study through a social media post.
The website requires the participant to agree to the participant information sheet and forwards them to the data collection page. On the data collection page, demographic data is surveyed from the participant and the hardware information is automatically detected.
Subsequently, the website sends the encrypted data to a University owned Qualtrics server using the Qualtrics \gls{api}. Subsequently, the participant obtains a payment for participating in the study.

\subsection{Acquisition of hardware components and benchmark}

The browser information and operating system information is obtained via the user-agent header that is sent with each http-request.
The WEBGL\_debug\_renderer\_info from the WebGL Web-API is used for obtaining the hardware specifications.
This data requires further processing as it includes the concatenated vendor information  such as the brand and model name, and the WebGL renderer.

An open benchmark dataset is used as a comparative metric for analysing the overall performance of a system.
Subsequently, a dataset that contains the benchmark results is obtained.
Datasets were searched via the google research dataset search utility\footnote{\href{https://datasetsearch.research.google.com/search?src=0&query=GPU\%20benchmark\%203D&docid=L2cvMTFyemRsY2M2dA\%3D\%3D}{Google dataset search} with query: "GPU benchmark 3D".} and a google search \footnote{\href{https://www.google.com/}{Google search} with the keywords: "GPU benchmark 3D download dataset" (first page only).}. 
Google database search did not yield any appropriate datasets for assessing 3D GPU performance.
The few results were either hardware specifications or mobile device benchmarks.
However, a Google search delivered eight datasets. 
These datasets were analysed based on (1) open research compliance and (2) covering all GPUs required based on obtained survey data \cite{paolacci2010running}.

\begin{table}[h]
    \centering
    \caption{Overview of considered datasets with criteria.}
    \begin{tabular}{l l l}
        \toprule
        Name & Open research & GPU Coverage\\
        \midrule
        3DMark.com & no & - \\
        PassMark & no & 200 (insufficient)\\
        GFXBench & no & 245 (insufficient)\\
        Blender Benchmark & yes, open & 248 (insufficient)\\
        UL Benchmarks & no & 199 (insufficient)\\
        UserBenchmarks & transparent and free & 1201 (sufficient)\\
        \bottomrule
    \end{tabular}
    \label{tab:datasetComp}
\end{table}

UserBenchmark's \cite{userbenchmark} dataset is used for evaluating the GPU performance, due to their transparent method for analysing hardware performance and sufficiently extensive collection size of benchmarks (compare to \autoref{tab:datasetComp}).
Their transparency qualities are a publicly available dataset and an open formula for calculating the benchmark metric. 
Their benchmarking score combines the average FPS with higher than average and lower than average fluctuations from multiple games (called \acrfull{efps}, see \autoref{eq:efps}) \cite{userbenchmark}.
\begin{align}
\begin{split}
EFPS = 0.35 * avg + 1.69/8 * (& \text{0.1\% lowest FPS average} \\ 
&+ \text{0.1\% lowest FPS max} \\
&+ \text{1\% lowest FPS average} \\
&+ \text{1\% lowest FPS max}) \label{eq:efps}
\end{split}
\end{align}


\subsection{Data analysis}

The hardware performance is reported in form of qualitative data but also quantitatively.
Qualitative information include used Operating systems, browsers, most commonly used GPUs that can qualify future research and policies.
These qualitative information are complemented by quantitative descriptive statistics and the evaluation of the hypothesis.
Both qualitative and quantitative information are synthesised to answer the research question.

R (v3.6.5) is used to test whether the data supports rejecting absence of hardware performance differences between \gls{mturk} and snowball recruiting. For this an appropriate test procedure and significance level must be defined. As convention, the significance level is set to $\alpha =0.05$.
In terms of testing procedure, one distinguishes between parametric tests that assume the validity of modeling the population distribution with a specific set of parameters and non-parametric tests that do not make such assumptions.
Whether the distribution is sufficiently modeled using \textit{normal distribution} parameters is tested with the Shapiro-Wilk test, and the necessary homoscedasticity is tested using the Levene's test.
If the data is both normal and homoscedastic, the Mann-Whitney U-test is used to evaluate the hypothesis. Otherwise, Wilcoxon's Rank-Sum test is used.


\section{Results}

\begin{figure}[h]
    \centering
    \includegraphics[width=\linewidth]{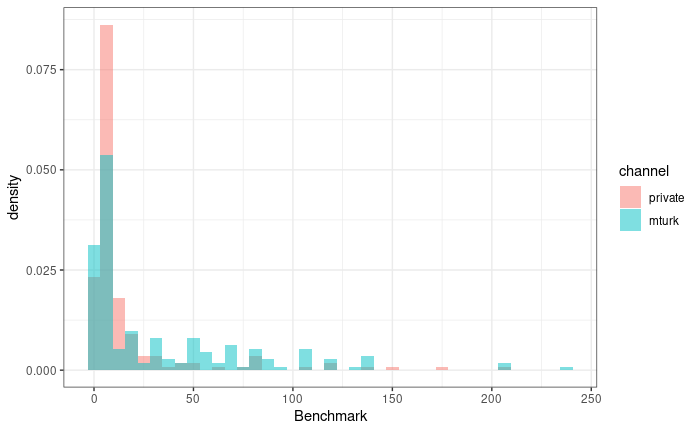}
    \caption{The relative frequency distribution of \gls{efps} (compare to \cite{userbenchmark}) per treatment group. The benchmark score density is plotted layered for \acrfull{mturk} and snowball recruitment (private). The bin-width is set to 6.25 as calculated by the Freedman–Diaconis rule.}
    \label{fig:benchmarkHisto}
\end{figure}

Firstly, the data is analysed in terms of distribution and relevant descriptive statistics. The dataset contains the $357$ samples from \gls{mturk} (n=179) and snowball (n=178) recruiting. The benchmark score's distribution spans between $0.65$ and $236$ \gls{efps} and indicates that most participants (68.0\%) reach a performance of less than 12.5 \gls{efps} (compare to \autoref{fig:benchmarkHisto}). The mean of the score is 24.24 and the median is 5.93. The overall operating systems running the browser were Windows 10 (88.0\%), Windows 8 (1.7\%), Windows 7 (3.4\%), MacOS 10 (4.8\%) and MacOS 11 (2.2\%). Additionally, the browsers running the web applications were Chrome (75.4\%), Firefox (24.4\%), and Safari (0.3\%).

\begin{figure}
    \centering
    \includegraphics[width=\linewidth]{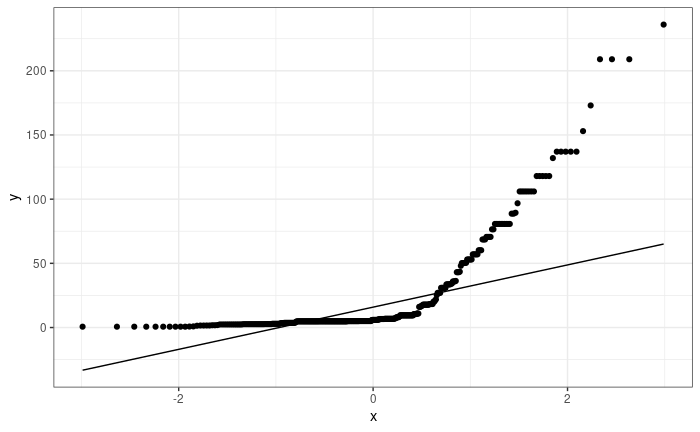}
    \caption{The QQ-plot of the data plot the normal distribution against the Benchmark dataset scores. The visual analysis suggests non-normal distribution because the data is not homogeneous with the normal distribution. }
    \label{fig:qq_plot}
\end{figure}

Secondly, the hypothesis tests are executed. Levene's test indicates homoscedasticity, $F=17.466$, $p<0.001$. The Shapiro-Wilk test indicated that the data is non-normally distributed, $W(356)=0.615$, $p<0.001$. This is confirmed by visual analysis (compare to \autoref{fig:qq_plot}). Therefore, the Wilcoxon's test is used to assess the hypothesis. The test indicates that the hardware performance of \gls{mturk} participants ($Mdn=7.16$) is significantly different to those recruited using snowballing ($Mdn=5.03$), $W = 19078$, $p=0.001$.

\section{Discussion and conclusion}

The data collection was limited by the certain available hardware information through Web-API. Thus, only allowing the hardware performance investigation based on the main graphical processing unit (GPU). While this provides a good estimate of the expected system's capabilities, a computer system's performance is influenced by more than the GPU \cite{gries2004methods}. Especially the interplay between CPU and GPU can significantly reduce the performance that would be possible with the GPU \cite{paolacci2010running}. Future research could investigate more advanced concepts such as inferring CPU models based on statistical modelling. Further research could analyse study-specific interactions of collectable data (such as the interaction of browser, graphics engine and GPU) on performance. However, since the used benchmark covers multiple configurations these interaction effects are assumed to be at lest mitigated.

Moreover, while the crowdsourcing data is expected to be adequately representative because of the large sample size \cite{mohan2017crowdsourcing}, the snowball recruitment was sampled in a Euro-centric and academia over-representative social network. Thus, results are expected to vary when the samples are drawn from different populations. Nevertheless, is the data sampling representative for many other European researchers and the methodology, novel data, and considerations reported are still integral for researchers to ensure quality and validity of experiments.

The results validate rejection of the null hypothesis in favor of the alternative that hardware performance in crowdsourcing and traditional (snowball) recruiting differs significantly ($p=0.01$). Indicating that applications have significantly different hardware conditions on \gls{mturk} than on traditional recruiting. 
Since the hardware on \gls{mturk} is significantly different, it might be necessary to filter participants or adapt the application to the crowdsourcing environment to obtain valid results \cite{aguinis2021mturk}.

However, the data also shows that there are vast intra-group differences. Both \gls{mturk} and traditional recruiting data is not normally distributed and has a high variance of performance scores. This distribution suggests that the application's performance shouldn't be tailored based on the average hardware performance of the either population.
The high spread of performances suggests the average performance is an insufficient metrics because the frames per second of lower than average performing hardware can still negatively affect experimental results. 
Many of the reported GPUs have \gls{efps} that would severely impede the user experience (under 30 FPS) \cite{chen2007review}. Thus, it would be more appropriate to tailor the application to a performance baseline that allows decent performance (e.g. above 30 FPS) on all targeted devices.

Ensuring appropriate application performance is possible by increasing application efficiency or by increasing the threshold of hardware requirements participants must suffice. Determining the application baseline is now possible based on the reported data, by investigating the distribution of devices with their respective \gls{efps} scores. If it is not possible to lower the application's requirements sufficiently by increasing software efficiency, externalising rendering to a server can ensure valid research results, given sufficient bandwidth to participants.


\bibliographystyle{IEEEtran}
\bibliography{Evaluating_hardware_differences_for_crowdsourcing_and_traditional_recruiting_methods}

\end{document}